\documentclass[12pt,twoside]{article}   
\usepackage[super,sort,comma]{natbib}

\usepackage{fancyhdr}		

\usepackage[section]{placeins}   %

\usepackage{graphicx}
\usepackage{multirow}
\usepackage{subfigure}

\makeatletter \renewcommand\@biblabel[1]{$^{#1}$} \makeatother
\newcommand{\cen}[1]{\begin{center} #1 \end{center}}

\usepackage[mathlines]{lineno}

\usepackage{hyperref}
\hypersetup{ colorlinks,
    citecolor=blue,
    filecolor=blue,
    linkcolor=blue,
    urlcolor=blue
}

\usepackage{xcolor}

\definecolor{gray}{rgb}{0.6,0.6,0.6}
\definecolor{red}{rgb}{0.85,0,0}
\definecolor{green}{rgb}{0,0.85,0}
\definecolor{blue}{rgb}{0,0,0.85}
\definecolor{beige}{rgb}{0.92,0.87,0.78}
\usepackage[all]{hypcap}    

\begin{document}

\cen{\sf {\Large {\bfseries A Unified Generation-Registration Framework for Improved MR-based CT Synthesis in Proton Therapy} \\  
\vspace*{15mm}
Xia Li\textsuperscript{1,2}, Renato Bellotti\textsuperscript{1,3}, Barbara Bachtiary\textsuperscript{1}, Jan Hrbacek\textsuperscript{1}, Damien C. Weber\textsuperscript{1,4,5}, Antony J. Lomax\textsuperscript{1,3}, Joachim M. Buhmann\textsuperscript{2}, Ye Zhang\textsuperscript{1}}\\
1. Center for Proton Therapy, Paul Scherrer Institut, 5232 Villigen PSI \\
2. Department of Computer Science, ETH Zürich, 8092 Zürich \\
3. Department of Physics, ETH Zürich, 8092 Zürich \\
4. Department of Radiation Oncology, University Hospital of Zürich, 8091 Zürich \\
5. Department of Radiation Oncology, Inselspital, Bern University Hospital, University of Bern, 3010 Bern
\vspace{5mm}
Version typeset \today\\
\vspace{5mm}
}

\pagenumbering{roman}
\setcounter{page}{1}
\pagestyle{plain}
\sf Author to whom correspondence should be addressed. Email: ye.zhang@psi.ch \\

\begin{abstract}
\noindent {\bf Background:} The use of Magnetic Resonance (MR) imaging for proton therapy treatment planning is gaining attention as a highly effective method for guidance. At the core of this approach is the generation of Computed Tomography (CT) images from MR scans. However, the critical issue in this process is accurately aligning the MR and CT images, a task that becomes particularly challenging in frequently moving body areas, such as the head-and-neck. Misalignments in these images can result in blurred synthetic CT (sCT) images, adversely affecting the precision and effectiveness of the treatment planning. 

\noindent {\bf Purpose:} This study introduces a novel network that cohesively unifies image generation and registration processes to enhance the quality and anatomical fidelity of sCTs derived from better-aligned MR images.

\noindent {\bf Methods:} The approach synergizes a generation network (G) with a deformable registration network (R), optimizing them jointly in MR-to-CT synthesis. This goal is achieved by alternately minimizing the discrepancies between the generated/registered CT images and their corresponding reference CT counterparts. The generation network employs a UNet architecture, while the registration network leverages an implicit neural representation of the Deformable Vector Fields (DVFs). We validated this method on a dataset comprising 60 Head-and-Neck patients, reserving 12 cases for holdout testing.

\noindent {\bf Results:} Compared to the baseline Pix2Pix method with MAE 124.95±30.74 HU, the proposed technique demonstrated 80.98±7.55 HU. The unified translation-registration network produced sharper and more anatomically congruent outputs, showing superior efficacy in converting MR images to sCTs. Additionally, from a dosimetric perspective, the plan recalculated on the resulting sCTs resulted in a remarkably reduced discrepancy to the reference proton plans.

\noindent {\bf Conclusions:} This study conclusively demonstrates that a holistic MR-based CT synthesis approach, integrating both image-to-image translation and deformable registration, significantly improves the precision and quality of sCT generation, particularly for the challenging body area with varied anatomic changes between corresponding MR and CT. 

\end{abstract}

\newpage     

\tableofcontents

\newpage

\setlength{\baselineskip}{0.7cm}      

\pagenumbering{arabic}
\setcounter{page}{1}
\pagestyle{fancy}
\section{Introduction}

Proton therapy is distinguished by its precision and minimal radiation exposure to non-target tissues, necessitating accurate anatomical information for effective dose delivery\cite{roadmap}. A critical aspect of this process requires the integration of Magnetic Resonance (MR) and Computed Tomography (CT) imaging\cite{kashani2018magnetic}. Although useful, due to the statistical global similarity metrics, cross-modality registration between MR and CT is often plagued by residual alignment issues, leading to inaccuracies in treatment planning. MR-based CT synthesis emerges as a pivotal solution in this context, particularly for Daily Adaptive Proton Therapy\cite{dapt} (DAPT), where reducing the daily radiation dose for imaging arises as a primary concern. By generating CT-like images from MR scans, this method bridges the gap between MR's superior soft tissue contrast and the essential electron density information provided by CT\cite{mrpt}. This approach not only offers the potential to bypass the misalignment issues\cite{liu2021synthetic} inherent in conventional multi-modality registration methods but also reduces imaging radiation to patients\cite{han2017mr}. The accuracy of these sCT images becomes especially crucial in proton therapy\cite{thummerer2020comparison}, where even marginal inaccuracies can significantly alter the proton dose distribution, ultimately affecting the treatment outcome. This challenge is further amplified in anatomical regions subject to movement or deformation, such as the head-and-neck area\cite{rigaud2015evaluation}.

The integration of deep learning (DL) technologies in the synthesis of CT images from MR data represents a notable advancement in medical imaging\cite{arabi2018comparative,chen2018u,bahrami2020new,maspero2018dose}. These DL methods utilize sophisticated neural networks, and adapt to complex feature recognition and learning, to effectively bridge the variations inherent between MR and CT imaging modalities\cite{emami2018generating}. Research has validated the efficacy of various MR and CT sequences in synthesizing CTs\cite{qi2020multi,tie2020pseudo,scholey2022generation,massa2020comparison,olin2020feasibility}, particularly those proficient in delineating bone structures\cite{kaushik2022region,wyatt2023comprehensive} or from low-field images\cite{cusumano2020deep,dinkla2019dosimetric}. Moreover, diverse training configurations and evaluation settings have been thoroughly investigated. These include different input dimensions\cite{dinkla2019dosimetric,fu2019deep} (2D, 2.5D, 3D), input sizes\cite{klages2020patch,patch-based,dinkla2019dosimetric} (patch-wise, full size), and imaging planes\cite{maspero2020deep,spadea2019deep,klages2020patch} (axial-only, three-planes). Additionally, several studies have focused on developing specialized network architectures for both generative\cite{olberg2019synthetic,liu2019mri,dinkla2018mr,xiang2018deep,zhao2023ct} and discriminative\cite{touati2021feature,sun2022synthesis} networks to enhance accuracy in CT generation, or compared the performance among different network structure designs\cite{fetty2020investigating,largent2019comparison,nie2018medical}. Innovative training losses\cite{kazemifar2019mri,liang2023bony}, data standardization\cite{andres2020dosimetry} and strategies\cite{reaungamornrat2022multimodal,pan20232d} have been devised to optimize model performance. Beyond image-level assessments, some approaches also integrate dosimetric-level evaluations, applicable in various therapeutic contexts like IMRT\cite{chen2018u,liu2020abdominal}, VMAT\cite{liu2020abdominal}, proton therapy\cite{florkow2020deep,wang2022toward,wang2022toward} and other modalities\cite{zhao2023saru}. There also exist some works that consider the practical problems such as data from multi-centers\cite{boni2020mr,bird2021multicentre}, the truncated MR\cite{compensation}, intestinal gas\cite{olberg2021abdominal}, MR-linac workflow\cite{yjmdh20121synthetic} and etc.

Despite significant advances thanks to DL technologies, a critical challenge in MR-based CT synthesis is the scarcity of paired and ``ideal-aligned" training MR-CT pairs. The accuracy and quality of synthetic CT (sCT) images rely heavily on the quality of these data pairs\cite{liu2021synthetic,hooshangnejad2023deepperfect,mitigating}, making it crucial for clinical application. While some works explored the CycleGAN architecture to enforce cycle consistency\cite{lei2019mri,wang2022development,brou2021improving,yang2020unsupervised} for this data-lacking issue, it tends to generate artifacts and is less accurate than Pix2Pix when roughly aligned paired data are available\cite{peng2020magnetic,klages2020patch,galapon2023feasibility}. The collection of such paired MR-CT images demands close temporal proximity between the scans of the same patient, significantly limiting the number of viable cases. Furthermore, achieving alignment is challenging due to the inherent differences in tissue representation between MRI and CT and sequential image acquisition by the two modalities. Often, solutions have tended to apply deformable image registration (DIR) between the planning CT and MR to minimize the unavoidable anatomic differences. However, only global metrics like mutual information (MI) can be used for similarity comparison, which may not sufficiently capture local image details, leading to suboptimal registration. Contemporary methods on deformable registration are shifting towards a dual approach\cite{mckenzie2020multimodality,han2022joint}: employing a generation network to map images to a uniform modality, followed by registration using more sensitive metrics such as mean absolute error (MAE). This approach highlights the interdependent nature of conditional image generation and cross-modality registration, creating a complex interplay akin to a chicken-and-egg scenario. 

Our research proposes an innovative solution that jointly addresses these two challenges, introducing a groundbreaking framework that seamlessly integrates image generation and registration. This approach jointly optimizes the generation of synthetic CTs (sCTs) based on MR images and aligns these MR images with the planning CT. Contrary to prior methods that utilized generation to aid registration\cite{mckenzie2020multimodality,han2022joint}, our approach achieves the inverse effect, with deformable registration enhancing image generation quality. The registration component, constructed using the latest Implicit Neural Representation concept\cite{sitzmann2020implicit}, maps 3D coordinates and case identifiers to corresponding Displacement Vector Fields (DVFs), thereby encoding the deformations of training cases\cite{inrdir}. This robust registration fitting significantly aids the generation network in training progressively more aligned MR-CT pairs, ensuring high anatomical fidelity in the synthesized images. This method is particularly advantageous in dynamic regions, where conventional DIR methods are challenged due to misalignments in MRI and CT images. Consequently, the produced sCT images are more accurately value-mapped and anatomically precise, making them highly suitable for the stringent radiotherapy requirements, particularly for proton therapy planning. 

Our approach was thoroughly evaluated at both the image and dosimetric levels, demonstrating high conformity of the synthetic CT to the input MR and significantly reducing voxel-wise errors. These robust experimental results substantiate two key assertions:
1) Joint registration significantly enhances the training of the generation network. The process of alternating optimization functions is similar to coordinate optimization, where both generation and registration errors are minimized in alternation. This alternation leads to progressively better alignment, allowing the generation network to concentrate predominantly on image intensity mapping rather than compensating for spatial deviations. 2) Joint registration also significantly improves the quality of the resulting synthetic CT images. Evaluations using imperfectly registered MR and CT images can be misleading, as errors can arise from both conditional image generation and the misalignment of the dataset. By facilitating better-aligned MR images that closely resemble actual CT scans, our proposed framework effectively minimizes errors attributable to spatial deviations, thus offering a more accurate assessment of synthetic CT quality.

\section{Methods}

\subsection{Generation Network}

\begin{figure}[t!]
\vspace{5mm}
\centering
\includegraphics[width=0.8\textwidth]{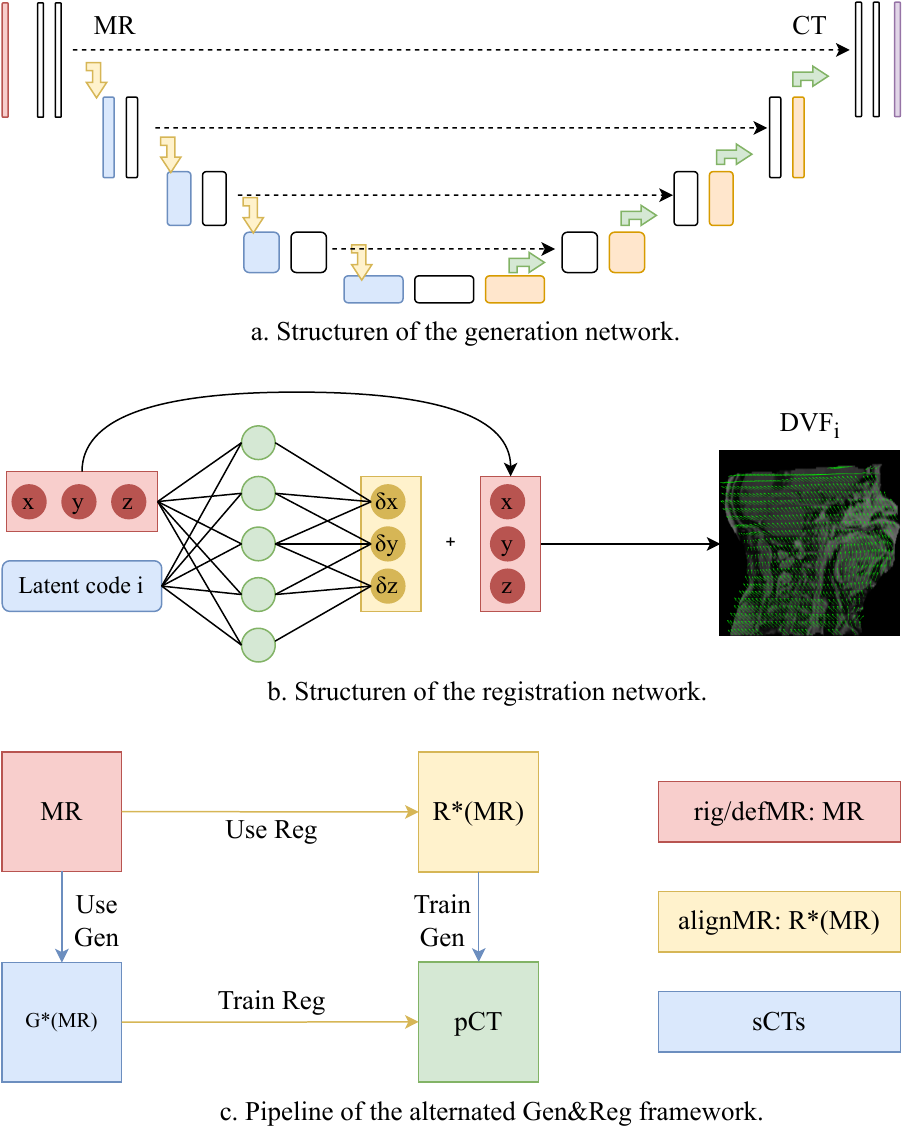}
\caption{Structures of the networks and pipeline of the alternated framework.\label{fig:structures}}  
\vspace{5mm}
\end{figure}

The UNet architecture employed in our study serves as the backbone for the MR-to-CT image generation process G, as shown in Fig.\ref{fig:structures}.a. This architecture is characterized by its symmetrical design, where the input MR images undergo a series of transformations through convolution, instance normalization, and max-pooling layers. These layers effectively compress the image into a condensed representation, capturing vital high-level features. Initially, the spatial dimensions of the image slice are mapped from $256 \times 256$ voxels to a compact feature map of $512\times 16\times 16$. Following this encoding pathway, the network transitions to a decoding phase. Here, deconvolution layers progressively restore the spatial dimensions, ultimately generating the sCT image. An essential aspect of the UNet architecture is implementing skip connections between corresponding encoding and decoding layers. These connections are crucial for retaining and incorporating intricate details from the MR images into the sCT output. For an in-depth understanding of the network's specific configurations and operational mechanisms, the readers are encouraged to consult the nn-UNet framework\cite{nnunet}.

\subsection{Deformable Registration Networks}

In image registration, conventional methods have predominantly focused on networks that learn to predict DVFs using paired inputs like MR and planning CT (pCT)\cite{cao2018deep}, or sCT and pCT\cite{kearney2018unsupervised}. These networks, trained on either 2D slices\cite{shan2017unsupervised} or low-resolution 3D images\cite{sentker2018gdl}, have limitations, especially in achieving high-quality 3D deformable registration. They are trained to generalize on testing datasets, instead of fitting the registration of training images.
Our method represents a departure from these conventional techniques. We have adopted an innovative approach to fit all DVFs for the training dataset. This strategy aligns more with our overarching goal of employing registration as an auxiliary tool in image generation. The implementation leverages the latest concept of Implicit Neural Representation (INR), an effective method for representing volumetric DVF's in continuous space\cite{idir} (IDIR). This technique maps the continuous input 3D coordinates $p_{j} = (x, y, x)$ of the $j$-th voxel to the corresponding displacement vectors $\delta p_{j} = (\delta x, \delta y, \delta z)$. Unlike the original IDIR \cite{idir}, where each case fits into a separate network, we use a shared Multilayer Perceptron (MLP), as illustrated in Fig.\ref{fig:structures}.b. Each case is identified by a unique learnable latent code $c_i$ as in \cite{dupont2022data}, streamlining the training process. This shared MLP model alleviates computational demands and considerably reduces memory and storage needs. The efficiency and scalability of our method are particularly beneficial when handling the large datasets commonly encountered in medical imaging.

\subsection{Alternated Optimization}

As depicted in Fig.\ref{fig:structures}.c, our novel approach alternatively optimizes the generation (G) and registration (R) networks through an alternating process. The entire framework was trained for $100$ epochs, with a $10$-epoch cycle marking the switch between G and R. This cycle commenced with G's training using the registered MR, denoted as R*(MR) to the pCT, followed by the generation of the sCT, represented as G*(MR) with the freshly trained G from the last iteration. Subsequently, R was trained using sCT as G*(MR) alongside the pCT. After this, DVFs inferred from R were applied to warp the MR into R*(MR), setting the stage for G's training in the subsequent cycle. After training, R*(MR) is referred to as \texttt{alignMR}, as it evolves from the original MR towards enhanced alignment with pCT. This aligned MR can be further utilized for more precise evaluations, as discussed in Sec.\ref{sec:metrics}.

Training of G involved an L1 loss calculation between R*(MR) and CT for each sampled slice, with a batch size of 64 and a learning rate of $2e-4$. The optimization of R incorporated two loss components. The first, denoted as $L_{reg}$, was the L1 loss calculated between the warped sCT and the pCT, expressed as $L_{reg} = \sum_i\sum_j L1(sCT_i(p_{j} + \delta p_{j}), pCT_i(p_{j}))$. The second component involved regularization over the determinant of the Jacobian matrix for DVF on a per-voxel basis, aiming to maintain values close to 1 to minimize distortion. R was trained with a batch size of 64 and a learning rate of $1e-4$, randomly sampling 64 voxels within the region of interest (RoI) for each case. All training and evaluation were executed on 2 NVIDIA RTX 4090 GPUs.

This alternating optimization strategy was instrumental in ensuring that G accurately captured the anatomical details of the MRs, while R effectively aligned these details with the corresponding CTs. Through iterative refinement of both networks, our framework achieved synergistic enhancements in both the anatomical accuracy of the synthetic CT images and their alignment with planning CTs.

\subsection{Dataset}
This study employed a dataset comprising images from patients with head-and-neck tumors, treated at PSI between 2017 and 2022. We collected paired MR-CT scans from $60$ patients, dividing them into training, validation, and testing sets with respective counts of $40$, $8$, and $12$. Each participant underwent at least one MRI and one CT scan during the pre-treatment planning phase. A significant challenge encountered was the notable anatomical differences between MR and CT scans, largely due to disparate scanning equipment and variations in patient neck positioning during the two scanning sessions. For CT imaging, we used a Siemens Sensation Open CT scanner at a tube voltage of 120kV, achieving image resolutions of $1\times 1\times 2$ mm. MR scans were obtained using a 1.5T Siemens Aera MR scanner, with a voxel size of $1\times 1\times 1$ mm. Primarily, the T1 vibe Dixon tra sequence was employed for network training purposes.
The study was carried out in strict adherence to ethical standards for human data research. Informed consent was obtained from all patients to use their anonymized data in scientific research. Anonymization was meticulously carried out before the analysis phase to ensure patient confidentiality and adherence to ethical standards.

\subsection{Preprocessing}
Both MR and CT images underwent preprocessing to ensure consistency and optimal quality for training and evaluation. Initially, we applied the Otsu method\cite{otsu}, using SimpleITK\footnote{\url{https://simpleitk.org/}}, to extract Regions of Interest (RoI) masks from each imaging modality. The final valid RoI region for analysis was determined by the intersection of these masks. In the final RoI, voxels outside this region were assigned a value of $-1024$ HU for CT and 0 for MR to maintain uniformity. MR images were normalized between $0$ and $1$, with extremities set at the $99.5$ and $0.5$ percentiles, while CT images were normalized within the Hounsfield unit range of $[-1024, 3072]$.

Subsequent steps involved aligning the MR images to their corresponding CT scans using mutual information-based registration methods in SimpleITK\footnotemark[1], resulting in two sets of images, \texttt{rigidMR} and \texttt{deformMR}, categorized by applying rigid or deformable registration. These images were resampled to match the CT resolution. For training our 2D generation network, slices were randomly sampled from these preprocessed images along the Z-axis. The training data were augmented per slice as in \cite{li2023uncertainty}, involving random rotations within a range of $[-30, 30]$ degrees, resizing factors between $0.7$ and $1.3$, and random flipping. During testing, MR slices were slices along the Z-axis without augmentation, and the output slices were stacked to reconstruct the synthetic CTs.

\subsection{Image-level Evaluation}\label{sec:metrics}
To ensure a thorough evaluation of the accuracy of our sCT images, we employed a diverse range of evaluation metrics, spanning two key aspects: intensity value mapping accuracy and anatomic conformity. For value mapping accuracy, we utilized Mean Absolute Error (MAE), Peak Signal-to-Noise Ratio (PSNR), and Structural Similarity Index (SSIM). MAE and PSNR calculations were confined to the RoI, while SSIM was calculated over the minimum-circumscribed cuboid of the RoI. To gauge anatomical conformity, we employed Dice Coefficients (Dice) for different regions (e.g. bone: HU $>250$) and Mutual Information (MI) between pairs of images. Specifically, MI served a dual role: it was used to evaluate the spatial invariance of the generation network by comparing MR and sCT images and to evaluate the efficacy of alignment by comparing different MR registration states \texttt{rigidMR}/\texttt{deformMR}/\texttt{alignMR} with the pCT. We plotted example MRs, pCTs, sCTs, and corresponding error maps to highlight the differences between different methods and evaluation MRs scenarios. In addition, 2D histograms between CT and different MRs were visualized to highlight spatial alignments.

For comparative analysis, we contrasted our proposed alternating methods, \underline{rigid$\to$align} and \underline{deform$\to$align}, against conventional sequential approaches using \texttt{rigidMR} (\underline{rigid-only}) or \texttt{deformMR} (\underline{deform-only}). These conventional methods involve training G with pre-processed registration between MR and pCT, while the new approaches signify an advance by incorporating the alignment process into the generation training. For evaluation, as the fidelity of pCT and MR was not assured, we employed both \texttt{deformMR} and \texttt{aligMR} as references with respect to the sCT. In particular, \texttt{alignMR} for the test set was generated by hold-out training on a reversed train/val/test split.

\subsection{Dosimetric-level Evaluation}

The ultimate aim of our work is to synthesize high-quality sCT for adaptive proton therapy. Therefore, dosimetric evaluations were performed on 12 cases with full information on the testing set. To assess the effectiveness of the image synthesis framework, we optimized the plan using the pCT and subsequently recalculated the doses for the sCTs of all scenarios using the Juliana software\cite{juliana}. This comparison entailed measuring the Mean Dose Error (MDE) at various dose thresholds (10\%, 50\%, 90\% of the maximum prescribed dose) between the doses optimized for pCT (pDose) and its recalculation for sCTs (sDoses). Additionally, we evaluated the gamma pass rate of pDose and sDoses using three distinct parameter sets (1mm/1\%, 2mm/2\%, 3mm/3\%). To facilitate a more intuitive comparison between pDose and sDoses, we plotted Dose-Volume Histograms (DVHs) of PTVs and OARs and extracted their representative dosimetric indices. Furthermore, our analysis also extends to present the differences in dose indices for both planning target volumes (PTVs) and organs at risk (OARs). 

\section{Results}
The proposed method exhibited good efficiency, with the training completed in only 1.5 hours, thanks to refined code optimization and the adoption of half-precision computation. The inference time for each test case was fast, taking only about 5 seconds per volume. The results presented in this section include the mean values and standard deviations across the $12$ testing cases to provide a comprehensive overview.

\subsection{Evaluation for sCT Generation}

\begin{table}[h!]
\centering
\caption{Quantitative comparisons between the proposed methods and conventional ones. The best results are in \textbf{bold} and the second best with \underline{underlines}. Arrows indicate the preference order of a measure. 
$\uparrow /\downarrow$ denotes higher / lower is better. \label{tab:gen}}
\vspace{3mm}
\footnotesize
\setlength\tabcolsep{7.4pt}
\begin{tabular}{c|c|ccc|cc}
\hline
Test                   & Train                   & MAE (HU) $\downarrow$               & PSNR (dB) $\uparrow$               & SSIM (\%) $\uparrow$               & Dice (\%) $\uparrow$               & MI (\%) $\uparrow$                 \\ \hline \hline
\multirow{4}{*}{\rotatebox{90}{\texttt{deformMR}}} 
                          & rigid-only       & 157.75±19.01        & 22.00±0.87          & 69.02±4.00          & 51.87±6.69          & 67.65±3.83          \\
                          & deform-only      & 124.95±30.74        & 23.78±1.53          & 73.81±4.44          & 67.78±12.60         & 72.26±4.69          \\
                          & rigid$\to$align  & 136.49±28.16        & 22.87±1.27          & 73.16±4.40          & 59.83±9.78          & 71.90±4.56          \\
                          & deform$\to$align & 115.65±33.87        & 24.31±1.77          & 76.42±4.86          & 68.14±12.66         & 74.55±5.01          \\ \hline
\multirow{4}{*}{\rotatebox{90}{\texttt{alignMR}}}  
& rigid-only       & 164.02±14.93             & 21.71±0.70                & 68.28±3.77              & 50.39±4.48            & 66.87±3.88          \\                          
& deform-only      & 111.36±8.86              & \underline{24.59±0.67}    & 74.60±2.95  & \underline{74.34±4.29}& 71.87±4.74          \\
& rigid$\to$align  & \underline{109.52±9.66}  & 24.51±0.68                & \underline{76.62±2.95}              & 68.37±3.44            & \underline{72.09±4.68}    \\
& deform$\to$align & \textbf{80.98±7.55}      & \textbf{27.13±0.85}       & \textbf{81.14±2.70}     & \textbf{79.48±3.70}   & \textbf{74.74±5.10} \\ \hline
\end{tabular}
\vspace{5mm}
\end{table}

\begin{figure}[h!]
\centering
\includegraphics[width=\textwidth]{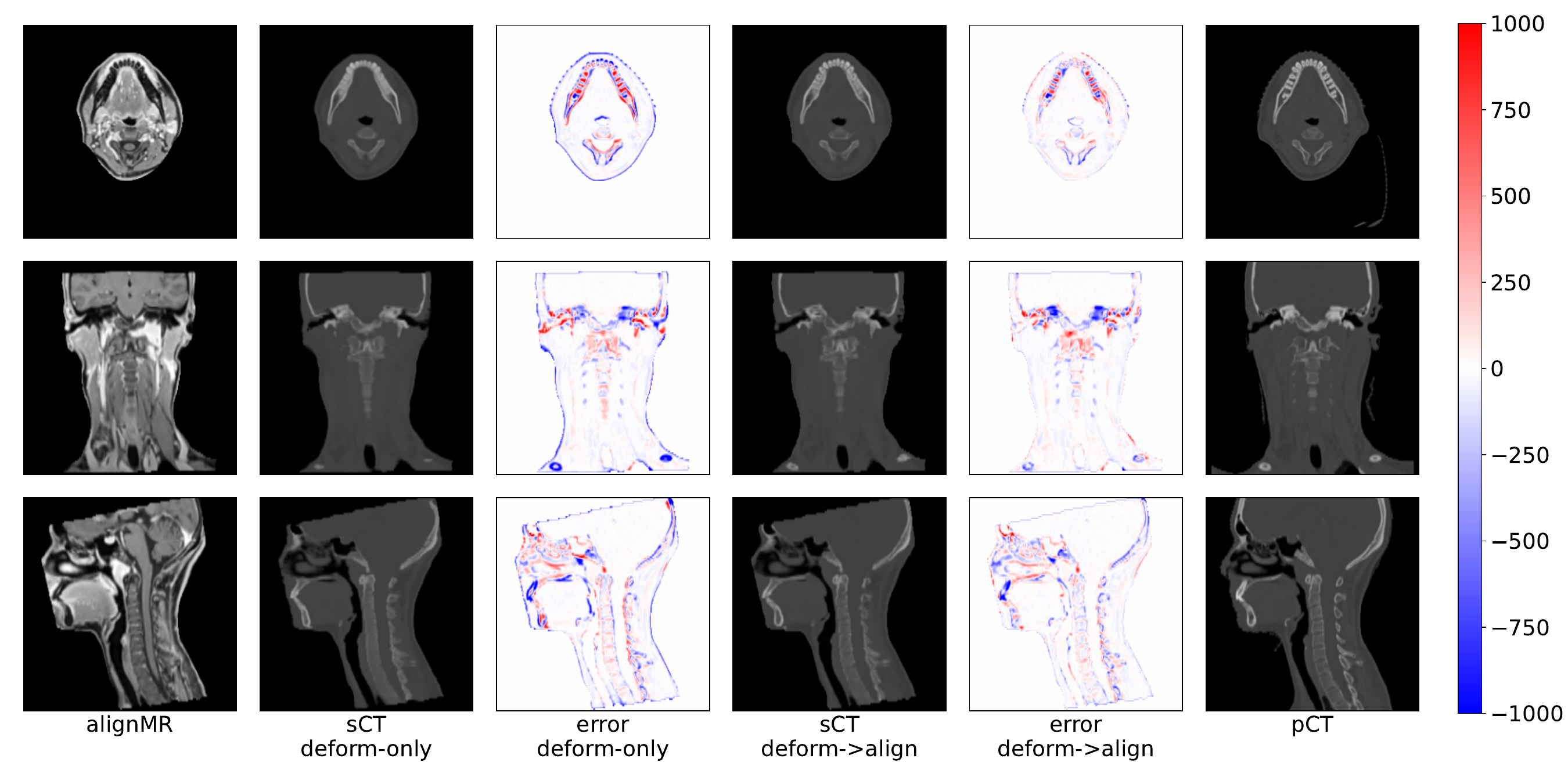}
\caption{Visual comparisons of sCTs and error maps from \underline{deform-only} and \underline{deform$\to$align} evaluated on \texttt{alignMR}. The error map was computed as sCT - pCT.\label{fig:vis-gen}}
\vspace{5mm}
\end{figure}

To assess the image generation performance, as a baseline reference, the conventional \underline{deform-only} method, when evaluated on \texttt{deformMR}, resulted in an MAE of $124.95 \pm 30.74$ HU and a Dice score of $67.78 \pm 12.60\%$ in the bone region across $12$ test cases. Our proposed \underline{deform$\to$align} method, evaluated on the same \texttt{deformMR}, demonstrated an improved MAE of $115.65 \pm 33.87$ HU and a marginally higher Dice score of $68.14 \pm 12.66\%$. Notably, the performance difference was more pronounced when tested on \texttt{alignMR}, which is considered as better aligned MR images to pCT. Here, the MAE for \underline{deform$\to$align} significantly reduced to $80.98 \pm 7.55$ HU, with the Dice score increasing to $79.48 \pm 3.70\%$. A similar pattern was observed in the comparison of \underline{rigid-only} with \underline{rigid$\to$align}, as detailed in Table.\ref{tab:gen}. In Fig.\ref{fig:vis-gen}, we visually compared sCTs generated by different methods with respect to \texttt{alignMR}, along with their corresponding error maps. More comprehensive results of MAE and Dice in different regions (air: HU $<-200$, tissue: $-200\leq$ HU $\leq250$, bone: HU $>250$) are presented in Tables~\ref{tab:mae}\&\ref{tab:dice} of the Appendices. Additional visual results are available in Figures 1\&2 of the Supplementary Materials.

\subsection{Alignment Evaluation}

\begin{figure}[h!]
\centering
\includegraphics[width=\textwidth]{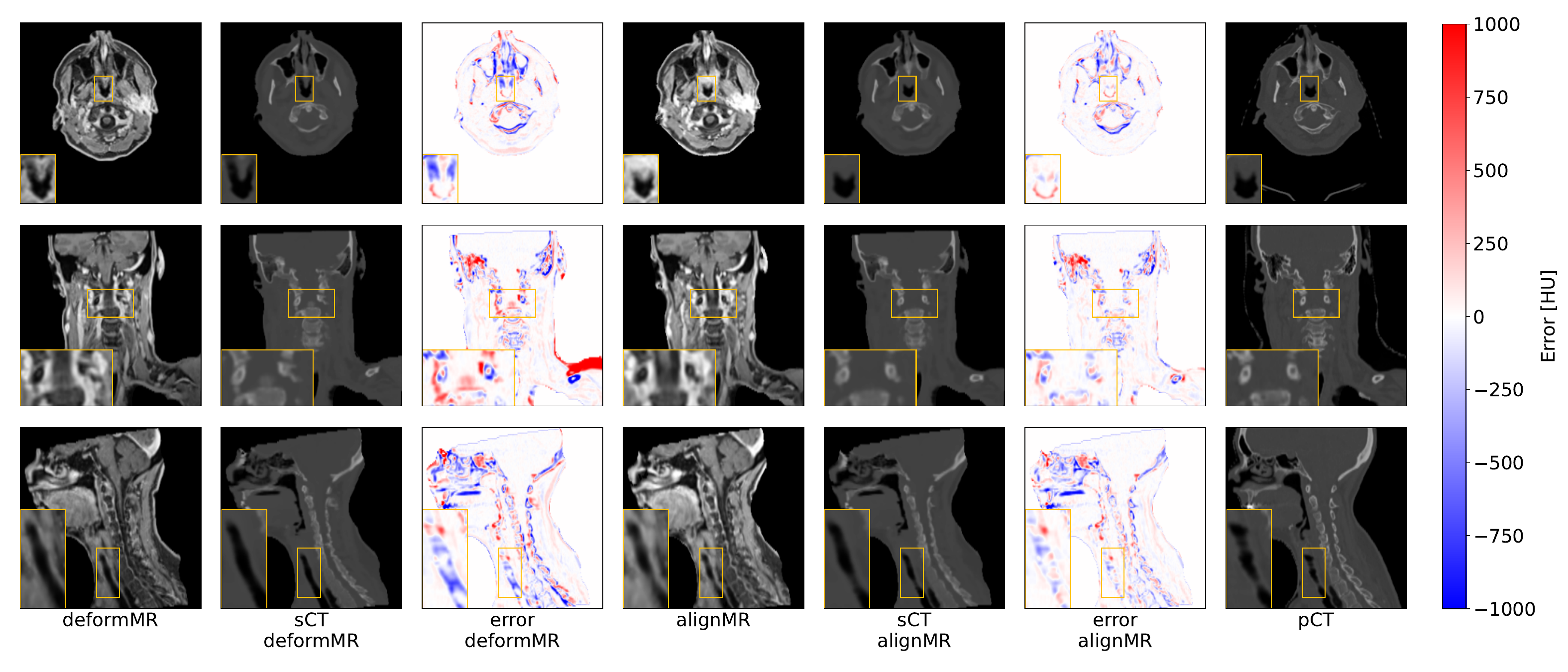}
\caption{Visual results of sCTs and error maps from \underline{deform$\to$align} on different MRs.\label{fig:vis-reg}}
\vspace{5mm}
\end{figure}

\begin{figure}[h!]
\centering
\vspace{3mm}
\includegraphics[width=0.8\textwidth]{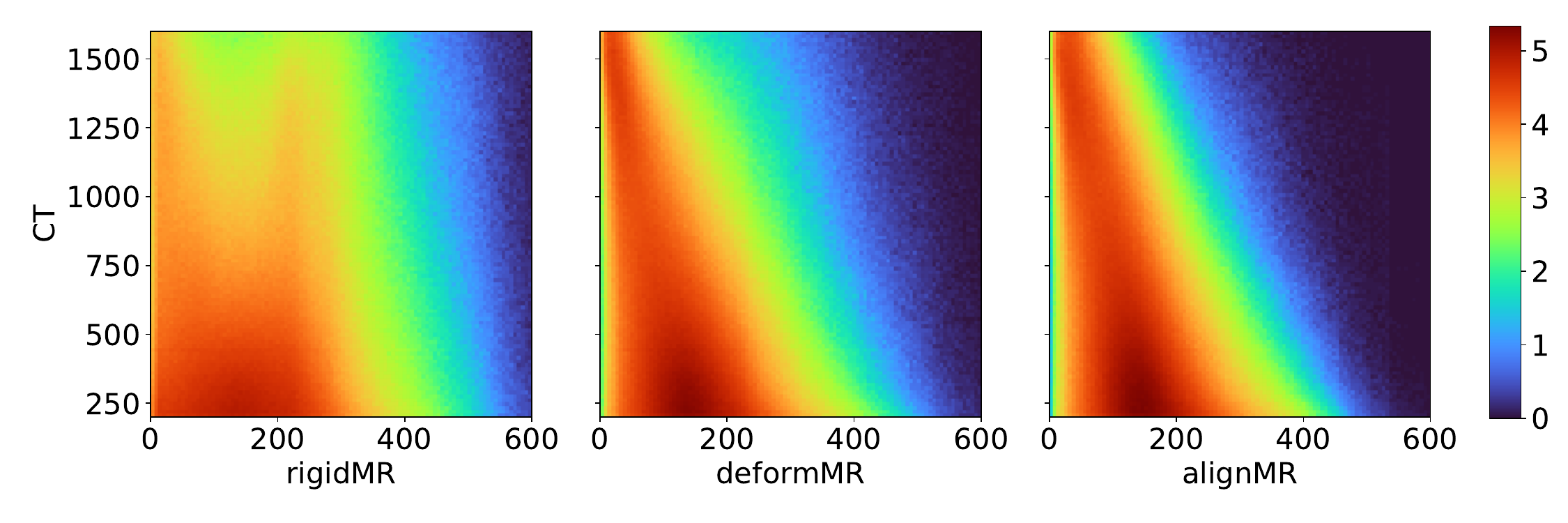}
\caption{Corresponding MR-to-CT log-scale joint histogram over bone regions. A narrower distribution corresponds to a higher MI value.\label{fig:hist}}
\vspace{5mm}
\end{figure}

The mutual information (MI) metrics further demonstrated the improvement of the paired images. The MI between input MR and pCT was $63.12 \pm 3.75\%$ for \texttt{rigidMR}, $67.76 \pm 4.46\%$ for \texttt{deformMR}, and significantly higher of $69.82 \pm 4.49\%$ for \texttt{alignMR}. This increasing trend in MI scores emphasizes the superior alignment achieved through our comprehensive registration approach. The refined alignment in \texttt{alignMR} led to fewer misalignments in the sCTs, allowing for a more precise comparison in terms of MAE and Dice scores as reported above. Fig.\ref{fig:vis-reg} displays sCTs generated by the \underline{deform$\to$align} method across various MR alignment states. Notably, \underline{rigid$\to$align} demonstrated comparable performance to \underline{deform-only} when evaluated on \texttt{alignMR}, indicating that a single training session on both generation and registration can match the performance of the conventional sequential deformable registration followed by generation training. Figure~\ref{fig:hist} presents joint histograms of MR-to-CT values, where \texttt{alignMR} exhibits a narrower distribution in the high CT value region compared to both \texttt{rigidMR} and \texttt{deformMR}.

\subsection{Dosimetric-level Evaluation}

\begin{table}[ht]
\centering
\caption{Gamma pass rate (\%) of sDoses from different methods over different thresholds.\label{tab:gamma}}
\vspace{3mm}
\footnotesize
\setlength\tabcolsep{18.0pt}
\begin{tabular}{c|c|ccc}
\hline
Testing                   & Training                   & \multicolumn{1}{c}{1mm/1\%} & \multicolumn{1}{c}{2mm/2\%} & \multicolumn{1}{c}{3mm/3\%} \\ \hline
\multirow{4}{*}{\texttt{deformMR}} & rigid-only                 & 91.13±5.71                  & 95.55±3.34                  & 97.50±2.09                  \\
                          & deform-only                & 92.73±5.10                  & 96.58±2.84                  & 98.17±1.69                  \\
                          & rigid$\to$align  & 93.40±5.09                  & 96.62±3.19                  & 97.99±2.27                  \\
                          & deform$\to$align & 93.83±4.94                  & 96.93±3.08                  & 98.21±2.24                  \\ \hline
\multirow{4}{*}{\texttt{alignMR}}  & rigid-only                 & 91.01±5.45                  & 95.60±3.16                  & 97.59±1.93                  \\
                          & deform-only                & 93.01±4.66                  & 96.99±2.41                  & 98.50±1.32                  \\
                          & rigid$\to$align  & \underline{95.04±3.79}            & \underline{97.79±1.90}            & \underline{98.86±1.05}            \\
                          & deform$\to$align & \textbf{95.85±3.31}         & \textbf{98.24±1.63}         & \textbf{99.11±0.94}         \\ \hline
\end{tabular}
\vspace{5mm}
\end{table}

\begin{figure}[ht]
\centering
\includegraphics[width=\textwidth]{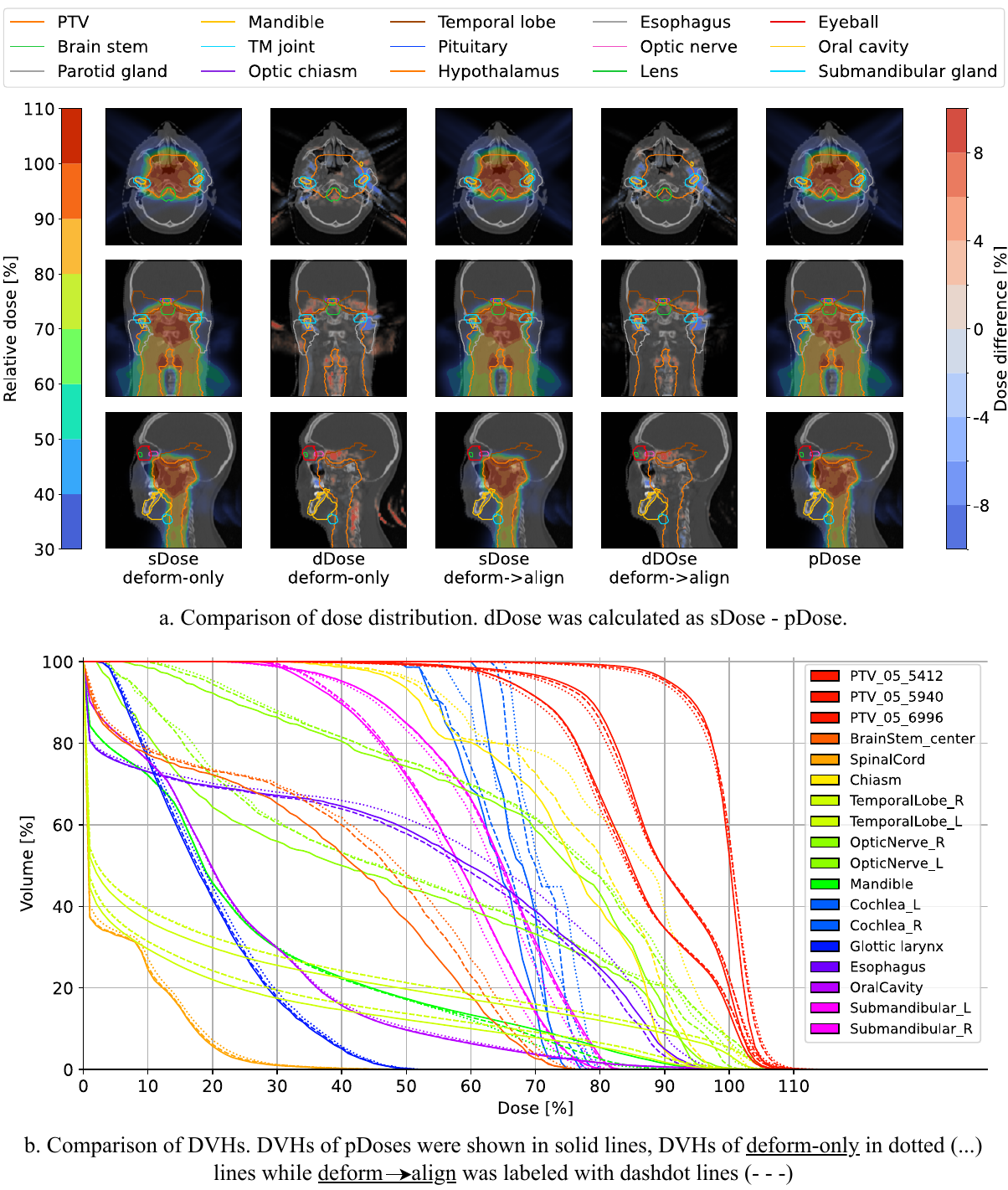}
\caption{Comparison of dose distribution and DVHs between conventional method \underline{deform-only} and 
 the proposed \underline{deform$\to$align}.\label{fig:dose-dvh}}
\vspace{5mm}
\end{figure}

\begin{figure}[ht]
\centering
\includegraphics[width=\textwidth]{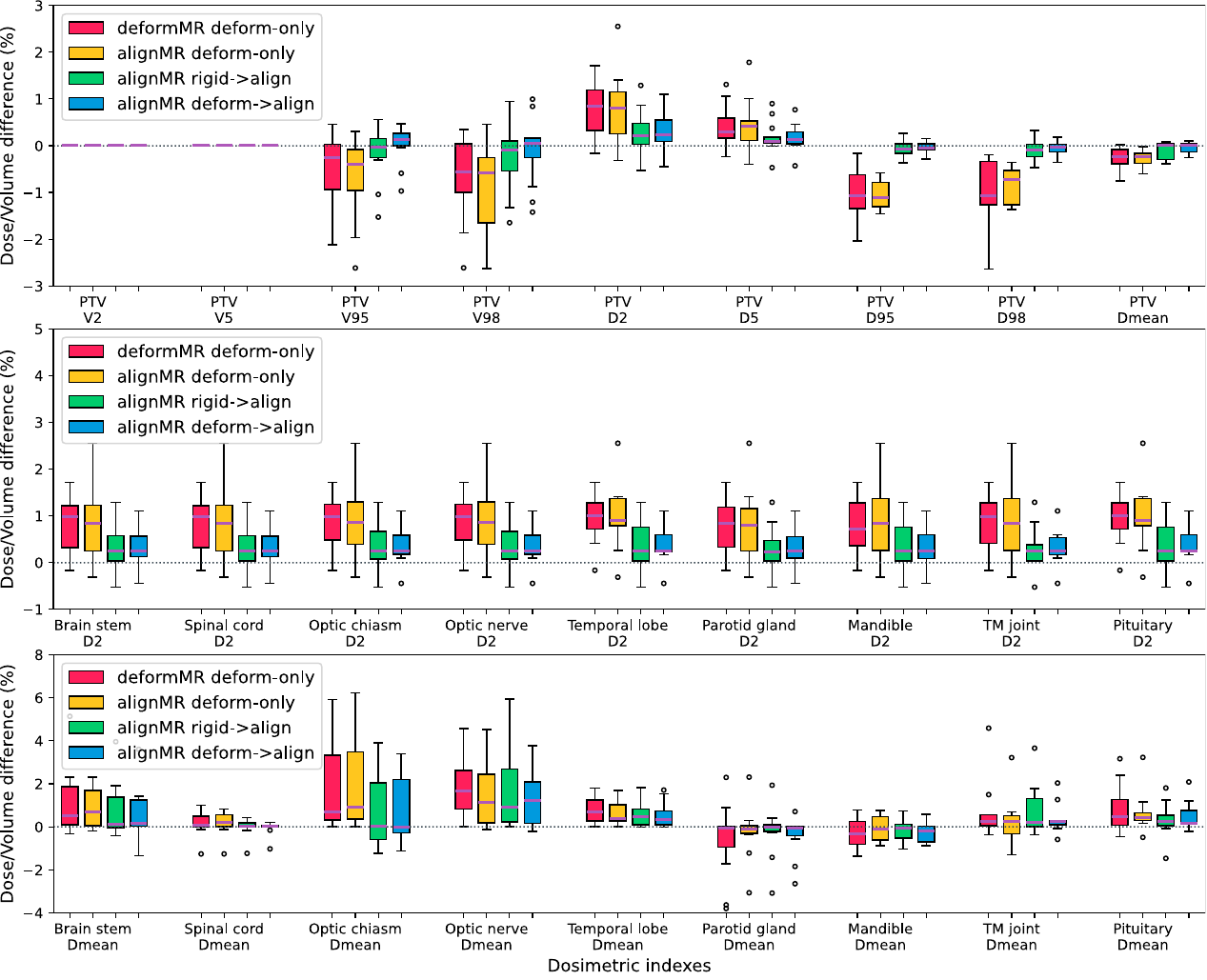}
\caption{Distributions of dosimetric index differences, compared across different settings. (Boxplot derived from the evaluations of all 12 testing patient cases).\label{fig:index}}
\vspace{5mm}
\end{figure}

Mean voxel-wise differences of the dose distribution on the body region calculated on the sCT generated by \underline{deform$\to$align} were less than $0.01\%$ of the prescribed dose when evaluated on \texttt{alignMR}. As a comparison, conventional \underline{deform-only} achieved a dose error rate of around $0.14\%$ on both \texttt{deformMR} and \texttt{alignMR}. More detailed results on dose error rates are shown in Table.~\ref{tab:dose}. With respect to the Gamma pass rate, the proposed method achieved $99.11 \pm 0.94$ given the threshold 3mm/3\%. Gamma pass rates for different, more strict tolerances are also shown in Table.~\ref{tab:gamma}. Visual comparisons of one example case on sDoses and dDoses (sDose - pDose) are shown in Fig.\ref{fig:dose-dvh}.a, between \underline{deform-only} and \underline{deform$\to$align}, with associated DVHs shown in Fig.\ref{fig:dose-dvh}.b. Furthermore, the differences of various dosimetric indexes for the PTV and OARs between sDose and pDose are plotted in Fig.\ref{fig:index}, where the four method scenarios were compared laterally. Visualization of more cases can be found in Fig 3\&4 in the Supporting Materials.

\section{Discussion}
\vspace{-5mm}

Our study evaluates the feasibility and advantages of an alternated image generation and registration framework for synthesizing CTs from MRI in the head-and-neck region for proton therapy. We addressed the complex inter-dependency between MR-based CT synthesis and MR-CT deformable registration. This approach is crucial for expanding MR-based CT synthesis regions where anatomy can deform considerably between the two imaging modalities. By solving the misalignment issue during generation training and evaluation, our method is particularly suited for challenging areas like the head-and-neck.

Registering MR and CT images poses considerable challenges, especially in deformation-prone areas like the head-and-neck region. These challenges are accentuated by inherent discrepancies in coordinates, resolution, and scanning position between MR and CT modalities. Furthermore, differences in texture and contrast between these scans add extra complexity. These factors not only complicate the multi-modality deformable registration process but also determine the training of generation networks. Techniques such as CycleGAN, useful for unpaired data, have performance limitations compared to Pix2Pix at a certain alignment level\cite{peng2020magnetic,klages2020patch,galapon2023feasibility}. Similarly, Pix2Pix, preferred for its fidelity in matched pairs, still struggles with pixel-to-pixel misalignments of the two images, leading to overly smooth inferred sCT or spatial biases when trained with isotropic or directional misalignments, respectively. Addressing these issues is essential, as accurate registration plays a pivotal role in the practical evaluation of generation networks, ensuring that the quality of sCT images is not compromised by alignment errors. Although some works using the paired images from the integrated MR-Linac system can inherently benefit from the more aligned dataset thanks to the dedicated fixation device between MR and CT imaging \cite{yjmdh20121synthetic}, this newly proposed framework would contribute greatly for those centers that have only out-room MR system.

Our methodology introduces a novel integrated approach, combining generation and registration networks into a cohesive framework. This unified translation-registration model improves the alignment between MR and CT images, thereby enhancing the anatomical accuracy of sCT generation. By alternatively optimizing G and R, our method addresses the challenges of conventional techniques that treat image generation and registration as separate sequential entities. This approach ensures that the generation network focuses more effectively on the nuances of intensity mapping, while the registration network aligns the images with high precision. This leads to the generated sCT images being more accurate in terms of electron density representation and superior anatomical conformity, both essential for precise radiotherapy planning.

Moroever, we strategically decided to align MR images with pCT images instead of the reverse. This choice was crucial for preserving the integrity and consistency of the training process. At early stages of training, sCT images have inferior quality and are often blurred; therefore, aligning them to fine-detailed MR images is safer than the opposite direction. The reversed choice, as employed in the study cited by\cite{mitigating}, presents a risk: fine details in pCT, such as streaks and spots, might be lost when registered to the smoother sCT, and this transformation is irreversible. When tested on our \texttt{alignMR}, the reverse process of \underline{deform$\Rightarrow$align} yielded worse MAE at $105.15 \pm 6.97$ HU, PSNR at $25.36 \pm 0.57$ dB, SSIM at $93.95 \pm 0.84\%$, bone Dice at $76.51 \pm 3.21\%$, thereby significantly underperforming compared to the results reported in the last row of Table~\ref{tab:gen}.

Our registration network adopts an INR approach\cite{idir} for \textbf{fitting} DVFs, a significant departure from all classic DL methods that typically rely on 2D registration networks\cite{kong2021breaking} or 3D low-resolution networks for \textbf{predicting} DVFs from paired inputs\cite{reaungamornrat2022multimodal}. By employing a shared network with unique latent vectors for each case, our method efficiently captures the required deformations while minimizing demands on memory and storage. With its compact 4MB network size, this approach offers a scalable and practical solution in medical imaging. Direct fitting of DVF volumes, though appearing to be feasible, tends to lack efficiency due to slow convergence and high demands on memory ($100$MB+). This efficiency from our choice is particularly advantageous in handling large medical image datasets, ensuring that our method is not only effective in capturing essential deformation information but also practical for real-world clinical applications.

Evaluating the quality of sCT images using non-aligned MR-CT pairs adds complexity to the analysis. The proposed approach can conflate errors from the value mapping process with those arising from alignment inaccuracies. By integrating registration with generation, our method minimizes such conflated errors, focusing more accurately on the quality of value mapping. However, potential convergence issues remain a concern, where the generation network might learn incorrect mappings along with spatial shifts, and the registration network could inadvertently correct these shifts, leading to low training errors but inaccurate predictions. Additionally, the use of MI as a primary metric for alignment evaluation has its limitations, as it does not fully capture the nuances of anatomical alignment, nor does it provide an explicit upper bound for registration quality. Furthermore, it remains challenging to ascertain the exact degree to which our \texttt{alignMR} could be considered an ``ideal" aligned MR, leaving room for further future exploration in this aspect of this method.

MR-based dose planning presents unique challenges due to the discrepancies in the Field of View (FOV) between MR and CT scans. Typically, out-room MR imaging focuses on a narrower FOV, primarily covering the Gross Tumor Volume (GTV) and its surroundings, whereas CT scans cover the full area, including the complete lymph nodes in the neck region. This difference poses challenges in ensuring comprehensive coverage in MR-based synthetic CT generation, which is critical for bridging the gap between experimental settings and practical clinical applications\cite{compensation}. In the dosimetric evaluation of this work, we chose to pad sCT to the full CT lending values from pCT, which may influence the performance measurements. Regarding the dose planning approach, we opted to optimize on pCT and recalculate on sCTs. Although the alternative approach of optimizing on sCT and recalculating on pCTs closely mirrors real-case scenarios where only MR is available, our choice aligns enough with our objective of effectively evaluating our synthesis framework, providing comparable data from both pCT and sCT for a more accurate assessment.

Finally, the definition of masks used in evaluation metrics like Mean Absolute Error (MAE) significantly influences the outcomes of our comparisons. Variations in mask definitions across different studies, such as using a strict body mask or an expanded mask to include more tissue types, make direct comparisons of MAE values challenging. This variability highlights the need for standardized evaluation protocols in medical imaging research. Such standardization would ensure fair and accurate comparisons between methodologies, particularly on public datasets with pre-defined metrics. Addressing these discrepancies is vital for advancing the field and enhancing research findings' reliability and clinical relevance. In this regard, SynthRAD\cite{synthrad} has made a significant step forward.
 
\section{Conclusion}
In this study, we developed an approach that substantially improves the anatomical accuracy of synthetic CT (sCT) images for MR-based radiotherapy planning. By integrating image generation with the deformable registration processes, the novel framework effectively addresses the inherent challenges in MR-CT registration and sCT evaluation. The methodological synergy of generation and registration networks, coupled with a strategic approach to proton dose planning, contributes to enhancing treatment planning accuracy. This study represents a meaningful step forward in the application of MR-based proton treatment planning and daily adaptive replanning, offering potential improvements in reducing the imaging-related dose and patient outcomes in the field of radiotherapy.

\clearpage

\section*{Acknowledgments}
\addcontentsline{toc}{section}{\numberline{}Acknowledgments}
This project is supported by the interdisciplinary doctoral grant (iDoc 2021-360) from the Personalized Health and Related Technologies (PHRT) of the ETH domain, Switzerland. The automatic planning system used in this work was developed as part of the EU-H2020 project `INSPIRE' (INfraStructure in Proton International REsearch; grant ID: 730983)

\section*{Appendix}
\addcontentsline{toc}{section}{\numberline{}Appendix}
\section*{More Quantitative Analysis}

\begin{table}[ht]
\centering
\caption{MAE values (HU) from different methods over different regions.\label{tab:mae}}
\vspace{3mm}
\footnotesize
\setlength\tabcolsep{9.6pt}
\begin{tabular}{c|c|cccc}
\hline
Testing                   & Training                   & \multicolumn{1}{c}{MAE\_body} & \multicolumn{1}{c}{MAE\_air} & \multicolumn{1}{c}{MAE\_tissue} & \multicolumn{1}{c}{MAE\_bone} \\ \hline
\multirow{4}{*}{\texttt{deformMR}} & rigid-only & 157.16±19.01            & 387.20±48.59             & 86.40±9.36               & 565.60±62.59             \\
                          & deform-only         & 124.85±30.74            & 329.58±74.43             & 71.69±17.24              & 412.87±93.90              \\
                          & rigid$\to$align     & 136.49±28.16            & 367.36±103.38            & 71.98±15.04              & 492.67±71.71             \\
                          & deform$\to$align    & 115.65±33.87            & 335.77±96.60             & 61.79±19.14              & 401.12±95.03             \\ \hline
\multirow{4}{*}{\texttt{alignMR}}  & rigid-only & 164.02±14.93            & 402.96±37.35             & 91.03±6.02               & 583.58±54.92             \\
                          & deform-only         & 111.36±8.86             & 306.05±40.63             & 64.38±3.77               & \underline{365.25±31.67}       \\
                          & rigid$\to$align     & \underline{109.52±9.66} & \underline{288.54±47.73} &\underline{54.40±4.24}    & 427.30±34.29             \\
                          & deform$\to$align    & \textbf{80.98±7.55}     & \textbf{251.89±39.61}    & \textbf{41.63±2.89}      & \textbf{290.21±24.53}    \\ \hline
\end{tabular}
\vspace{5mm}
\end{table}

\begin{table}[ht]
\centering
\caption{Dices scores (\%) from different methods over different regions.\label{tab:dice}}
\vspace{3mm}
\footnotesize
\setlength\tabcolsep{12.0pt}
\begin{tabular}{c|c|cccc}
\hline
Testing                   & Training            & \multicolumn{1}{c}{Dice\_mean}& \multicolumn{1}{c}{Dice\_air}& \multicolumn{1}{c}{Dice\_tissue}& \multicolumn{1}{c}{Dice\_bone} \\ \hline
\multirow{4}{*}{\texttt{deformMR}} & rigid-only & 65.02±3.40                    & 52.08±3.92                   & 91.11±1.21                      & 51.87±7.62                    \\
                          & deform-only         & 73.66±6.49                    & 59.90±6.10                   & 93.29±1.99                      & 67.78±12.60                   \\
                          & rigid$\to$align     & 71.21±7.08                    & 60.83±10.26                  & 92.97±1.90                      & 59.83±9.78                    \\
                          & deform$\to$align    & 75.77±7.85                    & 65.07±9.35                   & 94.08±2.18                      & 68.14±12.66                   \\ \hline
\multirow{4}{*}{\texttt{alignMR}} & rigid-only  & 63.17±2.28                    & 48.51±3.88                   & 90.62±0.80                      & 50.39±4.48                    \\
                          & deform-only         & 76.06±2.35                    & 59.75±4.48                   & 94.09±0.60                      & \underline{74.34±4.29}         \\
                          & rigid$\to$align     & \underline{78.19±2.08}        & \underline{71.59±4.39}       & \underline{94.63±0.54}          & 68.37±3.44                    \\
                          & deform$\to$align    & \textbf{83.85±2.05}           & \textbf{75.93±3.66}          & \textbf{96.14±0.56}             & \textbf{79.48±3.70}           \\ \hline
\end{tabular}
\vspace{5mm}
\end{table}

\begin{table}[ht]
\centering
\caption{Mean Dose Error (\%) from different methods by different thresholds.\label{tab:dose}}
\vspace{3mm}
\footnotesize
\setlength\tabcolsep{12.0pt}
\begin{tabular}{c|c|cccc}
\hline
Testing                   & Training                   & \multicolumn{1}{c}{Dose\_body} & \multicolumn{1}{c}{Dose$>$10\%} & \multicolumn{1}{c}{Dose$>$50\%} & \multicolumn{1}{c}{Dose$>$90\%} \\ \hline
\multirow{4}{*}{\texttt{deformMR}} & rigid-only                 & 0.18±0.19                     & 0.23±0.25                    & 0.09±0.67                       & -0.15±0.62                    \\
                          & deform-only                & 0.14±0.12                     & 0.15±0.19                    & 0.12±0.46                       & 0.04±0.36                     \\
                          & rigid$\to$align  & 0.02±0.12                     & -0.03±0.28                   & -0.11±0.53                      & -0.12±0.44                    \\
                          & deform$\to$align & 0.03±0.09                     & -0.02±0.25                   & -0.07±0.46                      & -0.06±0.39                    \\ \hline
\multirow{4}{*}{\texttt{alignMR}} & rigid-only                 & 0.18±0.17                     & 0.24±0.24                    & 0.10±0.63                       & -0.18±0.60                    \\
                          & deform-only                & 0.14±0.12                     & 0.20±0.17                    & 0.18±0.39                       & \underline{0.04±0.39}               \\
                          & rigid$\to$align  & \underline{0.00±0.09}               & \underline{-0.01±0.14}             & \underline{-0.05±0.36}                & -0.07±0.35                    \\
                          & deform$\to$align & \textbf{0.00±0.06}            & \textbf{-0.01±0.12}          & \textbf{-0.02±0.24}             & \textbf{-0.00±0.23}           \\ \hline
\end{tabular}
\vspace{10mm}
\end{table}

\section*{References}
\addcontentsline{toc}{section}{\numberline{}References}
\vspace*{-10mm}





\bibliography{./example}      



\bibliographystyle{./medphy.bst}    


\end{document}